# Strain-tunable magnetism and nodal loops in monolayer MnB

Chang Liu,[1,a)] Botao Fu,[2] Huabing Yin,[1] Guangbiao Zhang,[1] and Chao Dong[3,a)]

[1] *Institute for Computational Materials Science, School of Physics and Electronics, International Joint Research Laboratory of New Energy Materials and Devices of Henan Province, Henan University, Kaifeng 475004, China*

[2] *College of Physics and Electronic Engineering, Center for Computational Sciences, Sichuan Normal University, Chengdu 610068, China*

[3] *Institute of High Energy Physics, Chinese Academy of Sciences, Beijing 100049, China*

Designing two-dimensional (2D) materials with magnetic and topological properties has continuously attracted intense interest in fundamental science and potential applications. Here, on the basis of first-principles calculations, we predict the coexistence of antiferromagnetism and Dirac nodal loop (NL) in the monolayer MnB, where the band crossing points are very close to the Fermi Level. Remarkably, a moderate strain can induce an antiferromagnetic to ferromagnetic phase transition, driving the monolayer MnB to a ferromagnetic metal with Weyl NLs. Such a type of topological quantum phase transition has not been observed before. In addition, the symmetry-protected properties of the two types of NLs as well as the magnetic critical temperatures are investigated. The controllable magnetic and topological order in monolayer MnB offers a unique platform for exploring topological quantum phase transitions and realizing nanospintronic devices.

[a)] Author to whom correspondence should be addressed: cliu@vip.henu.edu.cn; dongchao@ihep.ac.cn

The discovery of graphene[1] opens the gate for the field of two-dimensional (2D) materials. In the past few years, more and more 2D materials have been extensively studied, such as phosphorene[2], hexagonal boron nitride[3], transition metal dichalcogenides[4] and MXenes[5]. From a practical point of view, 2D materials permit much easier control of their properties compared with 3D bulk materials, such as by means of strain[6], electric fields[7], doping, etc. The ultra-thin size, high mobility, and exotic physical properties that can be easily regulated, suggesting 2D materials the best candidates for making devices at the atomic scale[8,9].

In the spintronic applications of 2D materials, two types of characteristics are very important: magnetism and topological property, and both of them have individually attracted the constant attention of a large number of researchers [10-12]. Magnetic properties, especially ferromagnetism, endow 2D materials with the ability to store information. Since 2017, several van der Waals crystals have been experimentally synthesized and found to possess intrinsic ferromagnetism, including $Cr_2Ge_2Te_6$[13], $CrI_3$[14], and $VSe_2$[15], and a number of 2D ferromagnetic materials are predicted by theoretical calculations at the same time[10,16-18]. On the other hand, the non-trivial topological state near the Fermi level will bring exotic effects on the transport properties, which are crucial for data transfer in spintronic devices. 2D topological materials with protected band crossing points, including 0D nodal points, and 1D nodal loops [19-22], have been systematically studied. Several 2D nodal-loop materials have been proposed in theory [6–10], of which a few have been confirmed in recent experiments, such as monolayer $Cu_2Si$[23], and $CuSe$[24]. Very recently, it was discovered that Weyl nodal lines and ferromagnetism coexist in monolayer $GdAg_2$[25], opening new perspectives for exploring topological band structure in connection with magnetic order on 2D scale [26-28].

Form the above considerations, one can see that it is desired to explore 2D materials hosting both topological order and magnetic order. However, it is much more desired to effectively tune these intriguing properties by simple means such as applying external fields or strains. The realization of tunable topological and magnetic orders is particular interesting because of its significance in both fundamental physics investigation and potential topological device applications. Such a realization in a realistic 2D system has not been reported so far.

Here, we show that monolayer MnB, which was theoretically proposed in 2017[29], can serve as a good candidate to achieve both the above objectives. Using first-principles method, we find that the ground state of 2D MnB is an antiferromagnetic semimetal, which hosts a Dirac nodal line (NL) surrounding the Γ point in the Brillouin zone (BZ). Interestingly, a small strain can induce a phase transition that transforms the system into a ferromagnetic state with two Weyl NLs. Both the Dirac and Weyl NLs are protected by the mirror symmetry. We demonstrate the thermodynamic stability of MnB at room temperature via *ab-initio* molecular dynamics simulation. We also estimate the Curie Temperature ($T_C$) and Neel Temperature ($T_N$) under different correlation strengths by performing Monte Carlo (MC) simulations. Our findings identify monolayer MnB as a promising material platform for exploring the interplay between strain-tunable magnetic order and topological non-trivial band structure.

The first-principles calculations were performed using the Vienna *ab initio* Simulation Package (VASP)[30] based on the density functional theory (DFT). The pseudopotentials are generated using the projector augmented wave (PAW) method[31]. The energy cutoff was set to be 400 eV for the plane-wave basis, and the Monkhorst-Pack k mesh with a size of 14 × 14 × 1 was adopted for the BZ sampling. We used a vacuum layer of 15 Å to avoid the artificial interactions between periodic images. The convergence criterion was set to $10^{-6}$ eV for the self-consistent electronic minimization and 0.005 eV/Å for forces during the ionic relaxation. Considering the strong correlation interaction between Mn 3d electrons, we used the LSDA+U method[32] and varied parameter U between 2 and 5 eV to see the effects of the on-site Coulomb repulsion on the electronic structure and magnetism of manganese. In general, we expect the value of U to be around 4eV for the Mn 3d states since this value is appropriate for manganese oxides[33], so in this work we mainly focus on the U=4 case. The MC simulations are performed on 2D 30×30 square lattice with periodic boundary condition using the standard Metropolis algorithm[34].

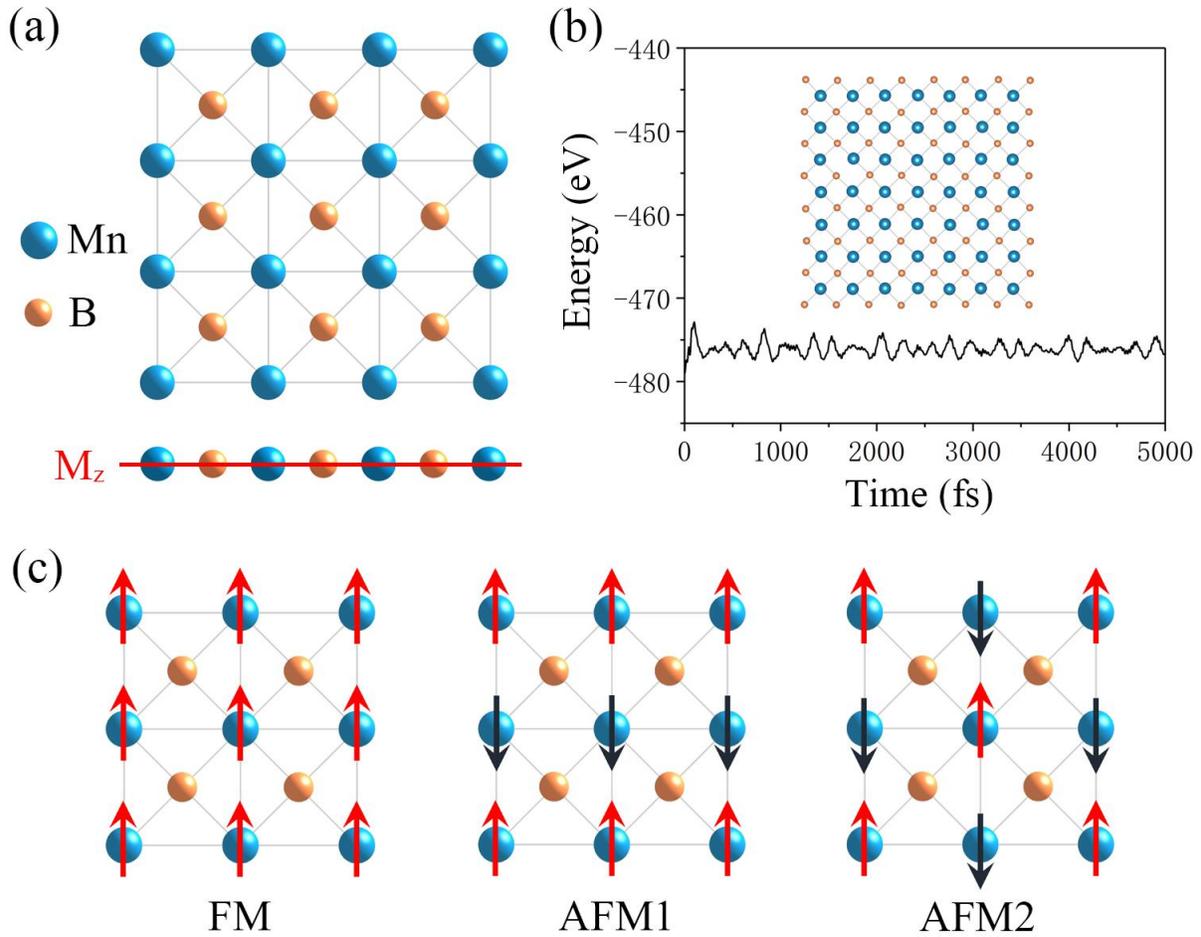

Fig. 1. (a) Top and side view of the lattice structure for monolayer MnB. The mirror plane ($M_z$) is indicated by the red line. (b) Time dependence of total energy in the AIMD simulation. The final geometric structure is shown in the inset. (c) Spin configurations of FM state, AFM1 state and AFM2 state, respectively.

As shown in Fig. 1 (a), the relaxed MnB lattice structure has a p4mmm symmetry with lattice constants a = b = 2.809 Å, and each of the Mn ions and B ions forms a tetragonal sublattice. The MnB monolayer has a purely flat structure, thus featuring a mirror reflection symmetry Mz. Our *ab-initio* molecular dynamic (AIMD) simulation result shows that freestanding single-layer MnB is stable at room temperature, as shown in Fig. 1 (b), after 5 ps MD simulation at 300K, both the supercell's total energy and the atomic positions have little fluctuation. The dynamical stability of monolayer MnB has also been confirmed by the calculation of phonon spectrum, as shown in Fig. S1 in the Supplementary materials (SM), there is no imaginary frequencies in Brillouin zones (BZ).

As shown in Fig. 1 (c), three different magnetic configurations are considered, including the ferromagnetic (FM) state, the type1 antiferromagnetic (AFM1) state where the spins are arranged anti-parallel along one of the two in-plane axes, and the checkboard-like type2 antiferromagnetic (AFM2) state. The LSDA+U calculations (U=4) show that the AFM2 state is the most stable, and its energy is about 0.3 eV (0.66 eV) per unit cell lower than that of the FM (AFM1) state. The magnetic moment of Mn ions is 3.46 $\mu_B$, 3.52 $\mu_B$ and 3.48 $\mu_B$ for the FM, AFM1 and AFM2 states, respectively.

Having confirmed the stability and identified the magnetic ground state of the monolayer MnB, then we investigated the detailed electronic properties. The band structure of MnB in AFM2 state is shown in Fig. 2 (a). We can see that the band structure shows a semimetal feature with the valence and conduction bands meeting in the vicinity of the Fermi level, forming three band crossings points marked by the red circles. Note that, although both the space inversion symmetry *P* and the time reversal symmetry *T* are broken by the magnetic order, the combined *PT* symmetry is preserved, so that each band is doubly degenerate. Therefore, all these linear crossing points show the Dirac features with four-fold degeneracy and locate around the Γ point, hinting at the emergence of the Dirac NL. The Heyd-Scuseria-Ernzerhof (HSE) hybrid functional[35] is used to check the band structure, as shown in Fig. S2 (a) in the SM, similar to the result of the LSDA+U method, three band crossings points also appear near the Fermi level. The semimetal feature can also be seen in the projected density of states (PDOS). The density of states at the Fermi level vanishes in a linear manner as in graphene, and the states near the Fermi level are mainly from the Mn-3d and B-2p orbitals. We further plot the 3D band dispersion to examine the Dirac feature in the whole BZ. As shown in Fig. 2 (b), one clearly observes that two bands near the Fermi level cross each other with opposite slopes, demonstrating a closed type-I NL surrounding the Γ point. Specifically, the linear dispersion range is large, and the NL has only a little dispersion, which should facilitate further experimental characterization.

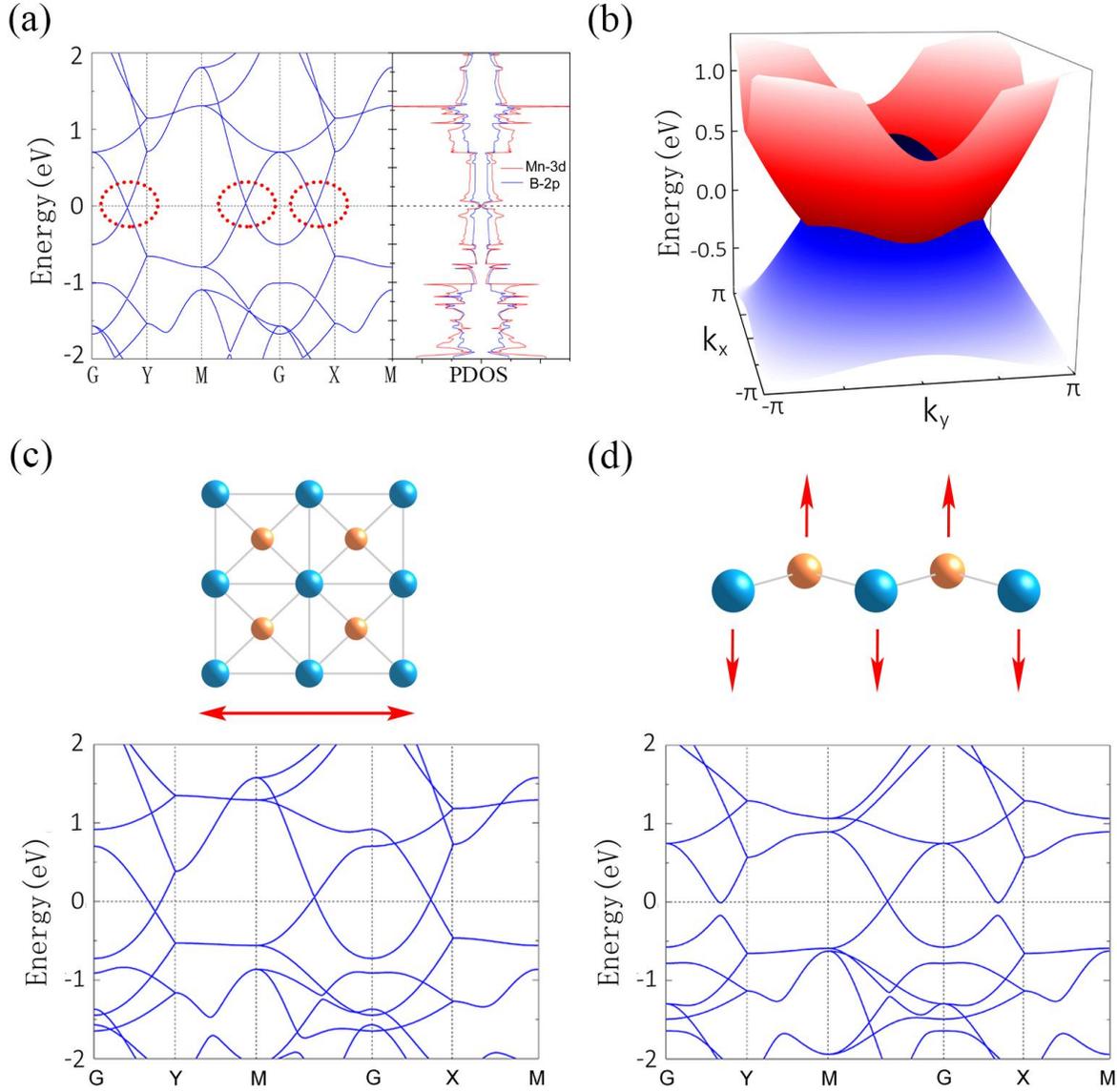

Fig. 2. (a) Band structure and PDOS of AFM2 MnB monolayer with U=4. (b) The 3D energy band dispersion. The band structure of monolayer MnB (c) under 5% uniaxial strain and (d) after breaking the $M_z$ symmetry.

In fact, the NL in monolayer MnB is protected by the mirror symmetry $M_z$. To demonstrate this, we performed two tests. First, we introduce a 5% uniaxial strain to break the $C_4$ rotation symmetry, as shown in Fig. 2 (c), the band structure shows little difference and the NL is preserved. In the second test, we artificially shift the positions of all the Mn atoms along the out-of-plane direction to break the mirror symmetry $M_z$. From the band structure in Fig. 2 (d), one observes that the crossing points at G-X line and G-Y line disappear, but the band crossing is still present at G-M line. Therefore, the system transforms into a Dirac semimetal. These results confirm that the NL is robust against in-plane strains and is protected by the $M_z$ symmetry.

Applying strain is a powerful method to modulate the physical properties of 2D materials because 2D materials usually have great mechanical flexibility[36-40]. By introducing biaxial strain to monolayer MnB, we find some interesting phenomena. As shown in Fig. 3, the total energy of different magnetic states can be tuned by strains. For the case of U=4eV, a 3% compressive strain will make the energy of the FM state lower than that of the AFM2 state, thus causing a magnetic phase transition. This phase transition preserves for U=2, 3 and 5 eV, but the required critical strains are different (1%~7% tensile strain). We further calculate the phonon spectrum of strained MnB (see Fig. S1). One can find that for both 3% compressive strain and 7% tensile strain, no imaginary frequencies appear in MnB, which indicates that MnB can withstand the strain that induces the magnetic phase transition. In addition, we notice that the correlation strength of Mn 3d electrons have an effect on the magnetism of MnB. When the correlation effect is relatively weak (U≤3), the FM state becomes the ground state for the unstrained structure (see Fig.3 (a-b)).

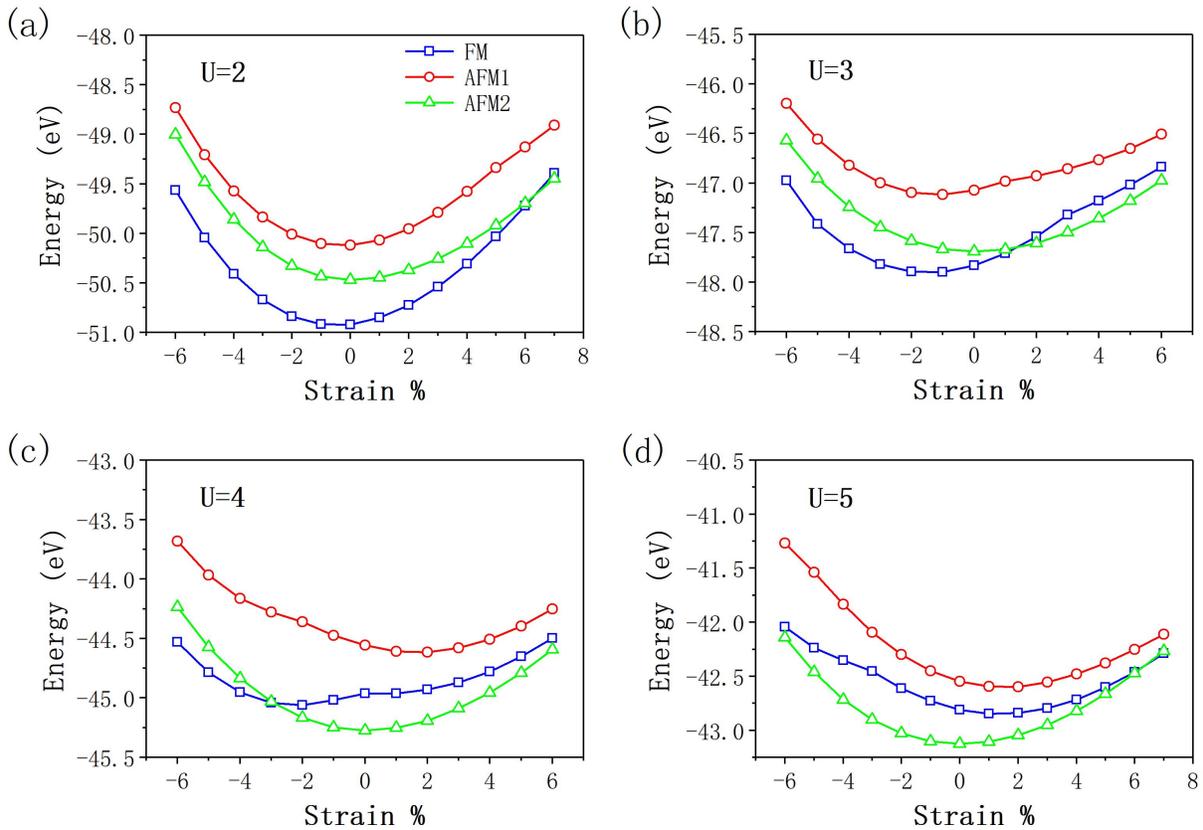

Fig. 3. The total energy of three magnetic states under different strains with effective on-site Coulomb interaction parameters (a) U=2 eV, (b) 3 eV, (c) 4 eV, and (d) 5 eV, respectively. The blue line with square, the red line with circle and the green line with triangle indicate the total energy of FM state, the AFM1 state and the AFM2 state, respectively.

Does the FM state also possess non-trivial topological electronic properties? In order to answer this question, we calculated the electronic structure of FM MnB (see Fig. 4 (a)). Above the Fermi level, there are two band crossing points on each path containing the Γ point (marked by the green circles); additionally, below the Fermi level, several other crossing points around the M point are observed (marked by the cyan circles). All the points are crossed by bands with opposite spin polarizations, forming two Weyl NLs in the BZ, surrounding the Γ point and the M point, respectively (see Fig. 4 (b)). All the crossing points persist in the HSE calculations (see Fig. S2 (b)), which confirms our result. We also broke the $C_4$ symmetry and the $M_z$ symmetry by artificially introducing strain and atom displacement. Different from the AFM2 state, all the crossing points remain gapless, so the NLs are preserved (see Fig. 4 (c-d)). This is because the spin is a good quantum number in the absence of spin-orbital coupling (SOC), and hence the spin degeneracy is intrinsic for monolayer MnB. As a result, the band crossing points in FM state cannot be gapped out by weak perturbations, even if the crystal symmetry is broken.

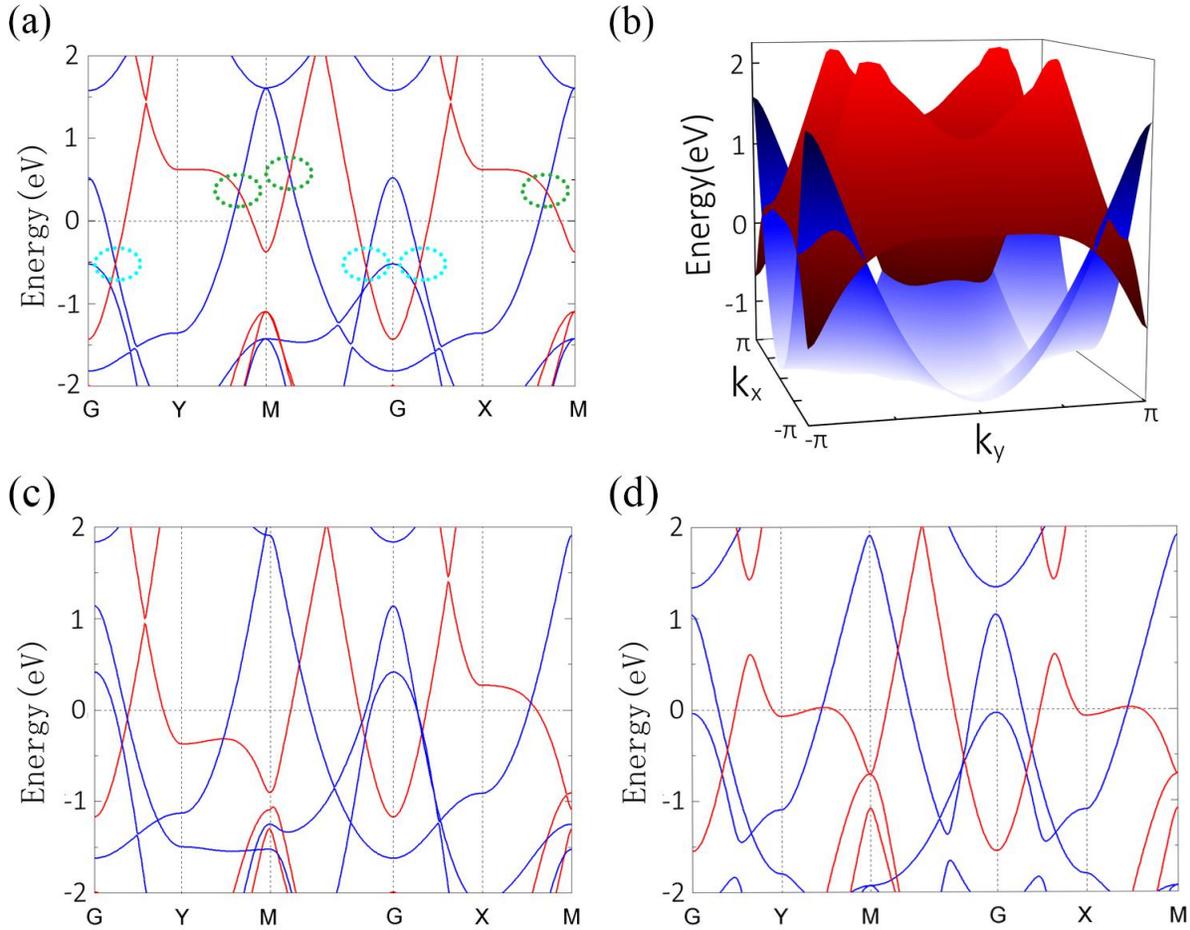

Fig. 4. (a) The band structure of FM MnB with U=4. (b) The 2D energy band dispersion. (c) The band structure of the FM state under 5% uniaxial strain. (d) The band structure of the FM state after breaking the Mz symmetry. The red line and blue line denote the bands with opposite spin polarizations.

When SOC is considered, the spin is no longer a good quantum number, and the $M_z$ symmetry is broken. Therefore, including SOC will gap out all the NLs both in the FM and AFM2 states. This conclusion is confirmed by our calculation with SOC (Fig. S3 in the SM). We find that most of the crossing points are gapped and hence the NLs are lifted. However, the SOC in MnB is rather weak, leading to a small band splitting. The opened energy gap (~22meV) is therefore negligible and the NLs are approximately preserved.

On the other hand, as we have mentioned before, the correlation strength (U) will influence the magnetic ground state, but it cannot change the symmetry-protected topological properties of MnB. For instance, we set the value of U to be 2, 3 and 5 respectively, and plot the corresponding band structures in Fig. S4. One finds that although the location of the band crossing points is altered, they remain gapless, preserving the NLs. Therefore, although the accurate value of U in monolayer MnB is hard to determine, we emphasize that both the strain-induced magnetic/topological phase transition and the topological properties are maintained within a large range of U values.

Next, we study the magnetic behavior in monolayer MnB under finite temperatures by performing classical Metropolis MC simulations in a Heisenberg's spin Hamiltonian: $\mathbf{H} = -\sum_{<ij>} J_{ij}\mathbf{S}_i \cdot \mathbf{S}_j - D(\mathbf{S}_{iz})^2$, where the summation $<ij>$ runs over all nearest-neighboring Mn sites, $J_{ij}$ is the exchange interaction strength between the normalized spin vectors $\mathbf{S}_i$ and $\mathbf{S}_j$. D is the single-site magnetic anisotropy parameter which can be obtained from the magnetic anisotropy energy ($E_{MAE} = E_{001} - E_{100}$), $\mathbf{S}_{iz}$ represents components of $\mathbf{S}$ along the $z$ orientation (out-of-plane), and $|\mathbf{S}| = \frac{3}{2}$ for Mn in 2D MnB because the moment of Mn atom is around 3 $\mu_B$. To obtain the unknown term $J_{ij}$, we first calculate the exchange energy $E_{ex}$ which difined as $E_{ex} = -2zJ_{ij}\mathbf{S}^2$, where z=4 is the the number of nearest-neighboring of Mn atoms. The $E_{ex}$ can be obtained from energy difference between the FM state and the AFM2 state ($E_{FM} - E_{AFM} = 2E_{ex}$)[41]. The calculated $J$ and $D$ from different spin configurations are listed in Tab.S1 in the Supplementary Materials.

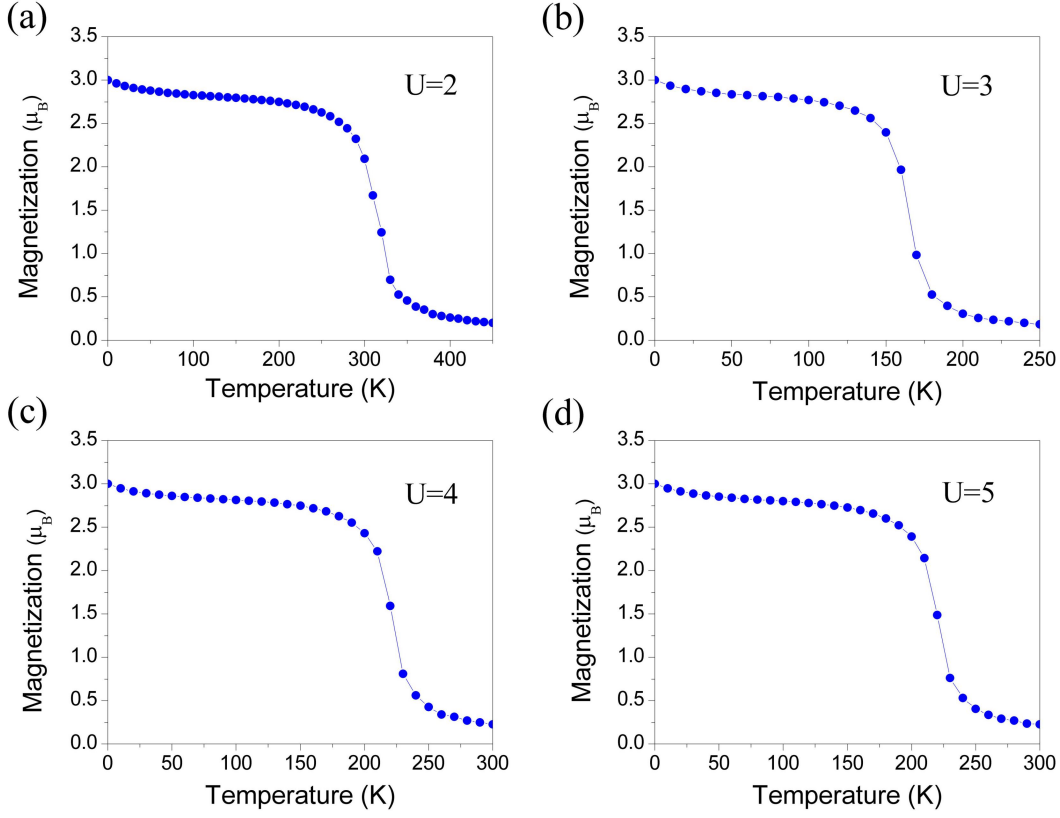

Fig. 5. Average sublattice magnetization as a function of temperature from MC simulations. Different correlation strengths with (a) U=2, (b) U=3, (c) U=4, and (d) U=5, have been considered.

The results of MC simulations are shown in Fig. 5. There are obvious differences between different U values. As mentioned earlier, the ground state for unstrained MnB is FM phase for U=2 and 3, and it turns to AFM state when U=4 and 5. For FM state, the $T_C$ is estimated to be ~325K for U=2, and ~175K for U=3, respectively. While the simulated $T_N$ (~225K) has little difference for U=4 and U=5. One observes that the enhanced correlation interaction significantly weakened the strength of ferromagnetic exchange whereas the anti-ferromagnetic coupling remains almost unchanged. The Ising model ($\mathbf{H} = -\sum_{<ij>} J_{ij} S_i S_j$) is also used to check the transition temperature. As shown in Fig. S5 in the SM, we can find that the critical temperatures for all the U values are slightly larger than the results of the Heisenberg model. This is because the magnetic anisotropy energy (MAE) in the Ising model is extremely large, and the large MAE in 2D materials will help stabilize the magnetism. Our results show that the magnetic critical temperature of MnB is higher than that of many discovered 2D magnetic materials such as $CrI_3$[13] and $Cr_2Ge_2Te_6$[14].

In summary, using first-principles method, we investigate the stability, magnetism, electronic structure and topological property of the monolayer MnB. We find that the correlation strength of Mn 3d electrons will affect the magnetic ground state. In particular, both the AFM2 and FM states have symmetry-protected topological states: the AFM2 state possesses a Dirac NL exactly at the Fermi level and the FM state exhibits two Weyl NLs. More importantly, these two phases can transform into each other under appropriate strains. The MC simulations results show that the magnetism can be stable near room temperature. The monolayer MnB thus offers an intriguing platform for the fundamental exploration of topological states and magnetic states. We believe that the results of this work are suitable for experimental verification and potential application.

**Supplementary Materials**

See supplementary materials for the phonon spectrum of freestanding and strained MnB in Fig. S1; band structure calculated by the HSE method in Fig. S2; band structure with SOC in Fig. S3; band structure with different U values in Fig. S4; $T_C$ and $T_N$ simulations with Ising model in Fig. S5; the calculated exchange parameters J and ionic anisotropic parameter D in Table S1.

This work was financially supported by the National Natural Science Foundation of China (Grant Nos. 11904079), the Strategic Priority Research Program of Chinese Academy of Sciences (XDB25000000) and the China Postdoctoral Science Foundation (No. 2019M652303).

**Data Availability Statements**

The data that supports the findings of this study are available within the article [and its supplementary material].